\documentclass[%
twocolumn,
reprint,
amsmath,amssymb,
aps
]{revtex4-1}

\usepackage{graphicx}
\usepackage{dcolumn}
\usepackage{bm}
\usepackage{hyperref}

\hypersetup{
	colorlinks=true,
	linkcolor=blue,
	filecolor=magenta,      
	urlcolor=blue,
	citecolor=blue,
	anchorcolor=blue,
}

\begin{document}
	
	
	\title{Theory of Polar Skyrmions in Layered Structure of Ferroelectric Perovskites}
	
	\author{Snehasish Sen}
	
	\author{Sudhansu S. Mandal}
	
	\affiliation{
		Department of Physics, Indian Institute of Technology, Kharagpur, West Bengal 721302, India
	}%
	
	%
	%
	
	\date{\today}

	\begin{abstract}
		Discovery of polar skyrmions on the (PbTiO$_3$)$_n$/(SrTiO$_3$)$_n$ superlattice in the absence of any known interaction, which can orient electric polarizations at a point with respect to its neighboring points, is enigmatic. We here show that the coupling of the electric polarization and strain at each layer is responsible for orienting polarizations. We formulate coupled Euler equations for electric polarization and elastic displacement vectors, and solve to find skyrmion solution for a range of electric fields with unique topological number in each layer. The types of skyrmions vary from Neel to Bloch depending on the position of the layers in the superlattice, as observed in the experiments. 
	\end{abstract}

	\maketitle
	

	Skyrmion in the condensed matter systems is a two-dimensional particle-like topological texture of a vector field having constant magnitude with an integral topological number.
	The vector fields for which skyrmions are realized are director fields with nonpolar head-tail symmetry in liquid crystals \cite{Smalyukh2010,Ackerman2014}, electronic spin magnetic moment in ferromagnetic quantum Hall systems \cite{Sondhi1993,Barrett1995}, magnetization in chiral magnets \cite{Muhlbauer2009,Yu2010}, electric field in optical guides \cite{Tsesses2018,Du2019}. Recent observation of skyrmions formed by electric polarization in  (PbTiO$_3$)$_n$/(SrTiO$_3$)$_n$ superlattice \cite{zhu2021dynamics,Das2019,Das2021} on the PbTiO$_3$ (PTO) planes is surprising because there is no {\em a priori} known interaction which will at least enforce the polarization vector at one position to tilt with respect to the same at a neighboring point.

	The possibility of forming polar skyrmions was first predicted in BaTiO$_3$ nano-wires embedded in a SrTiO$_3$ (STO) matrix \cite{Nahas2015} by a first principle calculation. An experimental investigation to this direction initiated on ferroelectric thin films finds the formation of bubble domains \cite{Zhang2017}. However, superlattice structures of ferroelectric materials are found to be promising for richer structures. Varieties of intricate polarization patterns have been identified, {\it e.g.}, vortex \cite{Yadav2016,Naumov2004,Hong2017,Chen2020,Abid2021,Li2021,Tan2021,Susarla2021}, skyrmion \cite{Das2019,Nahas2015,Goncalves2019,Nahas2020}, meron \cite{Nahas2020,Wang2020,Shao2023}, and hopfion\cite{Lukyanchuk2020}. In particular, the existence of polar skyrmion on the  (PTO)$_n$/(STO)$_n$ superlattice structures
	has been found in several experiments \cite{Das2019,zhu2021dynamics,Das2021}. However, the term `skyrmion' in this case should be considered with caution because the observed structure is in the form of a three dimensional bubble, in contrary to the skyrmions observed in chiral ferromagnetic systems as  two dimensional structures. Interestingly, within this bubble structure, two dimensional skyrmion structures have been observed in each layer of the PTO. The orientation of polarization changes from layer to layer making different types of skyrmions in different layers: Bloch type skyrmion in the central layer and a smooth variation of orientations lead to Neel type skyrmion in the top and bottom layers with inward and outward orientations respectively.

	While the three-dimensional phase-field simulations\cite{Das2019,Das2021,HONG2018155,zhu2021dynamics,Yuan2023,ZHOU2022,Hu2023} of free energy do produce bubble-like structures with a unique topological number\cite{Sk_number}, the microscopic mechanism for stabilizing such a structure in the absence of {\em a priori} known contribution in free energy that can orient electric polarization is not understood. In addition, the formation of different types of skyrmions in different layers of PTO is intriguing.

	In this letter, we unravel the hidden contribution in free energy that is responsible for essential linear in gradient of electric polarization contribution for the manifestation of difference in orientation of polarization at two neighboring spatial points.
	By forming and solving Euler equations for the polarization angle and the magnitude of elastic displacement vector causing strain, we find the stable skyrmion solution. Proposing an {\em ansatz} form of the displacement vector in different layers of PTO, we are able to find different skyrmion solutions in different layers, including Bloch type in the middle layer and Neel types in top and bottom layers, as reported in the experiments. We  estimate a range of effective electric field over which proliferation of stable skyrmions is possible. This range explains why skyrmion bubble is observed only for a range of layers ($12 \leq n\leq 18$) \cite{Hong2017,Junquera2023} in the superlattice (PTO)$_n$/(STO)$_n$.

	The free energy (excluding Landau free energy that stabilizes ferroelectric state) for a strained ferroelectric two-dimensional system with the application of electric field is given by \cite{Hu1998,LI2002} $F = \int_S dS[\mathcal{F}_{\rm gr} + \mathcal{F}_{\rm st} + \mathcal{F}_{\rm int}+\mathcal{F}_{\rm elec}]$. Here, the gradient surface energy density $\mathcal{F}_{\rm gr} =\frac{1}{2}g_{ijkl}P_{i,j}P_{k,l}$,  $(i,k =1,2,3 \,\, {\rm and} \,\, j,l=1,2)$, with $\bm{P}$ being electric polarization vector of unit magnitude, $P_{i,j} = \partial P_i /\partial x_j$ and the corresponding strength being $g_{ijkl}$ with underlying cubic lattice symmetry, {\it i.e.,} $g_{1111} = g_{2222}=g_{11}$, $g_{1122} =g_{2211}= g_{12}$, and $g_{1212}=g_{2121}=g_{3131}=g_{3232}= g_{44}$.
	$\mathcal{F}_{\rm st} =\frac{1}{2}c_{ijkl}\epsilon_{ij}\epsilon_{kl} $ is the contribution of surface density of strain energy, where strain tensor $\epsilon_{ij} = (u_{i,j}+u_{j,i})/2$ with $\bm{u}$ representing the elastic displacement vector of the defect causing strain, and the elastic stiffness tensor $c_{ijkl}$ satisfying relationship of symmetric stress tensor $\sigma_{ij} = c_{ijkl}\epsilon_{kl}$ with strain tensor possessing cubic lattice symmetry:  $c_{1111} = c_{2222}=c_{11}$, $c_{1122} =c_{2211}= c_{12}$, and $c_{1212}=c_{2121}=c_{3131}=c_{3232}= c_{44}$.
	The term $\mathcal{F}_{\rm int} = -q_{ijij}\epsilon_{ij}P_iP_j  $,  ($i\neq j$), represents interacting surface free energy density that couples electric polarization and strain with the corresponding coupling constant $q_{ijij} = q_{44}$. The other possible terms with $P_k^2$ (ignored here) contributes to the Landau free energy\cite{WANG2005} for the ferroelectric phase. Finally, $\mathcal{F}_{\rm elec}$ represents electrical free energy surface density
	where $\bm{E}$ is the net electric field, including the depolarizing field.

	Performing integration by parts in $\int_S dS\, \mathcal{F}_{\rm int}$, we obtain
	\begin{equation}
		\mathcal{F}_{\rm int} 
		\equiv  q_{44}  \biggl[  \sum_{i=1}^3\sum_{j\neq i, j=1}^2  \biggl( u_iP_i P_{j,j} +  u_iP_j P_{i,j} \biggr) 
		\biggr]
		\label{Fener_int}
	\end{equation}
	by neglecting boundary terms as we assume vanishing $\bm{u}$ at the boundary. Note that   $\mathcal{F}_{\rm int}$ in Eq.~\eqref{Fener_int} consists of linear gradients of polarizations. This will compete with $\mathcal{F}_{\rm gr}$ for relative orientation of polarization vectors in neighboring points.
	However, unlike key antisymmetric Dzyaloshinskii-Moriya interaction (DMI)\cite{DMI} with linear gradients of magnetization in chiral magnets for producing spin-spiral state and thereby magnetic skyrmions on the application of magnetic field, this interaction is not antisymmetric.

	The unit polarization vector may be parametrized by spherical polar variables $\Theta (\bm{{\rm r}})$ and $\Phi (\bm{{\rm r}})$ as $ \bm{P}(\bm{{\rm r}}) = \left[\sin\Theta(\bm{{\rm r}})\cos\Phi(\bm{{\rm r}}), \sin\Theta(\bm{{\rm r}})\sin\Phi(\bm{{\rm r}}), \cos\Theta(\bm{{\rm r}})\right]$. We consider the displacement vector (without loss of generality) in a system of $n^{\rm th}$ two-dimensional layer as
	\begin{equation}
		\bm{u}(r,\phi,z) \approx \bm{u}_{n}(r,\phi) = u_n(r) \left[ \cos\theta_{n} \hat{z}+\sin \theta_{n} \hat{r} \right]\, .
	\end{equation}
	Here $\bm{u}_n$ is considered isotropic in a layer and its dependence on different layers is attributed to the position of the layer through polar angle $\theta_n$ and  isotropic radial dependency of its magnitude $u_n(r)$. Considering the origin is at the central layer, $\cos(\theta_n) \approx 1$ at the top layer, $\cos(\theta_n) \approx -1$ at the bottom layer and $\cos(\theta_n) = 0$ at the central layer. The direction of $\bm{u}_n$ here is consistent with the experimentally found direction\cite{Das2019} of the displacement of Ti ion in different unit cells around the defect center.

	
	We consider $\Theta (r)$ has only radial dependence and $\Phi (\phi)$ has only angular dependence for searching a possible rotationally symmetric skyrmionic solution. Therefore in polar form \cite{suppli}, $F= \int_0^\infty dr\, r\int_0^{2\pi}d\phi\,  (\mathcal{F}_{\rm gr} + \mathcal{F}_{\rm elec} +\mathcal{F}_{\rm st} + \mathcal{F}_{\rm int} )$ with  $\mathcal{F}_{\rm gr} = \frac{g_{11}}{2}(\Theta_{r}^{2}+\frac{\sin^{2}\Theta}{r^{2}}\Phi_\phi^2)$, $\mathcal{F}_{\rm elec} = -E_3 \cos \Theta$, 
	
	\begin{widetext}
		\begin{eqnarray}
			\mathcal{F}_{\rm st} &=&\sin^{2}\theta_{n} \left[ c_{11} \frac{u_{n}^{2}}{2r^{2}}+c_{12}\frac{u_{n} }{r}\left( \frac{d u_n}{d r}\right)\right]-\frac{1}{4}\left[ \cos^{2}\theta_{n} (c_{11}+c_{12}) -2c_{11} \right] \left( \frac{ d u_{n}}{d r} \right)^2  \, ,\\
			\mathcal{F}_{\rm int} &=& \frac{q_{44}}{4} \, u_n \biggl[ \sin (\theta_n) \cos [2(\Phi-\phi)] 
			\left(\sin 2\Theta \, \Theta_{r} + \dfrac{2}{r}\sin^{2}\Theta \Phi_\phi \right) 
			+ \,4\cos (\theta_n) \cos(\Phi-\phi)\left(\cos 2\Theta\,\Theta_{r}+\dfrac{\sin 2\Theta}{2r} \Phi_\phi \right)\nonumber \\
			&& -\sin(\theta_n)\cos\left[2(\Phi+\phi)\right]\left( \sin 2\Theta \,\Theta_r -\dfrac{2}{r}\sin^{2}\Theta \Phi_\phi\right)     
			\biggr] \, .		 \label{Energy_int}   
		\end{eqnarray}
	\end{widetext}
	where $\Phi_\phi =d\Phi /d\phi$, $\Theta_r = d\Theta/dr$ and $E_3$  is the effective electric field along perpendicular to the plane.  For an isotropic environment, gradient energy coefficients \cite{WANG20137591} $g_{12} =0$ and $g_{11} = 2g_{44}$, and constraint relation \cite{STAAB199917} for elastic constants is $c_{11}-c_{12}=2c_{44}$.  Note that we work in a unit system in which electric polarization has been considered as unity.

	Since $\Phi(\phi) = \phi +\eta$ and $\Phi(\phi) = -\phi +\eta$, ($\eta$ being a constant to be determined by minimizing $F$), are necessary for describing \cite{Bogdanov1989,Bogdanov1994,Bera2019} a skyrmion and an antiskyrmion respectively, the first two terms in Eq.~\eqref{Energy_int} are in favor of skyrmions, and the last term is suitable for an antiskyrmion. For searching a skyrmion (antiskyrmion) solution, we therefore drop the last term (first two terms) in Eq.~\eqref{Energy_int} for constructing the corresponding Euler equation. This is allowed because skyrmions and antiskyrmions are topologically distinct. However, $F$ is determined by considering all the three terms in Eq.~\eqref{Energy_int} and the  skyrmion and antiskyrmion solutions will have vanishing contribution to the left-out term in free energy.
	The Euler equations for $\tilde{u}_n$ and $\Theta$ given by  $\frac{d}{dr}\left(\frac{\partial F}{\partial X_{r}}\right) - \frac{\partial F}{\partial X} = 0 $,    ($X=\tilde{u}_n \,\text{and} \,\Theta, $ $X_r = \partial X/\partial r$), are then found \cite{suppli} to be 
	\begin{widetext}
		\begin{eqnarray}
			&&	\left[2-\cos^{2}\theta_{n}\left( 1+\frac{c_{12}}{c_{11}} \right) \right]\left(\dfrac{d^{2}\tilde{u}_n}{d\rho^{2}}+\dfrac{1}{\rho}\dfrac{d\tilde{u}_n}{d \rho}\right)-2\sin^{2}\theta_{n}\dfrac{\tilde{u}_n}{\rho^{2}} = \dfrac{E_{0}^{\prime }}{E_{3}}\left[\sin \theta_{n} \cos (2\eta) 
			\left(\frac{\sin^{2}\Theta}{\rho}+\frac{\sin (2\Theta)}{2}\Theta_{\rho}\right)  \right. \nonumber \\
			&& \hspace{9cm}\left.+ \cos \theta_{n}\cos \eta\left(2\cos (2\Theta) \, \Theta_{\rho}+\frac{\sin (2\Theta)}{\rho}\right)\right]  \label{u_n}\\
			&&	\dfrac{d^{2}\Theta}{d \rho^{2}}+\dfrac{1}{\rho}\dfrac{d \Theta}{d \rho}-\dfrac{\sin 2\Theta}{2\rho^{2}} =-\dfrac{E_{0}}{E_{3}} 
			\biggl[ \dfrac{d \tilde{u}_n}{d \rho}  \cos 2\Theta\cos \eta\cos \theta_{n} 
			+\dfrac{1}{4} \cos 2\eta \sin 2 \Theta \sin \theta_{n}\left(\dfrac{ d \tilde{u}_n}{d \rho}-\dfrac{\tilde{u}_n}{\rho}\right) -\sin \Theta\biggr] \label{Theta}
		\end{eqnarray}
	\end{widetext}
	for a possible skyrmion solution,
	where dimensionless radial distance $\rho = r/r_0$ with $r_0 =U_0 q_{44}/E_3$, electric field scales $E_0 =  U_0^2q_{44}^2/g_{11}$  and $E'_0 = q_{44}^2/c_{11}$, and $\tilde{u}_n$ is in the unit of $U_0 = u_n (\rho =0)$. These two electric field scales are assumed to be same, {\it i.e.},  $E_0 \approx E'_0$. This identification fixes the value of $U_0$ as $U_0 = \sqrt{g_{11}/c_{11}}$.
	
	\begin{figure}[h]
		\centering
		\includegraphics[width=\linewidth]{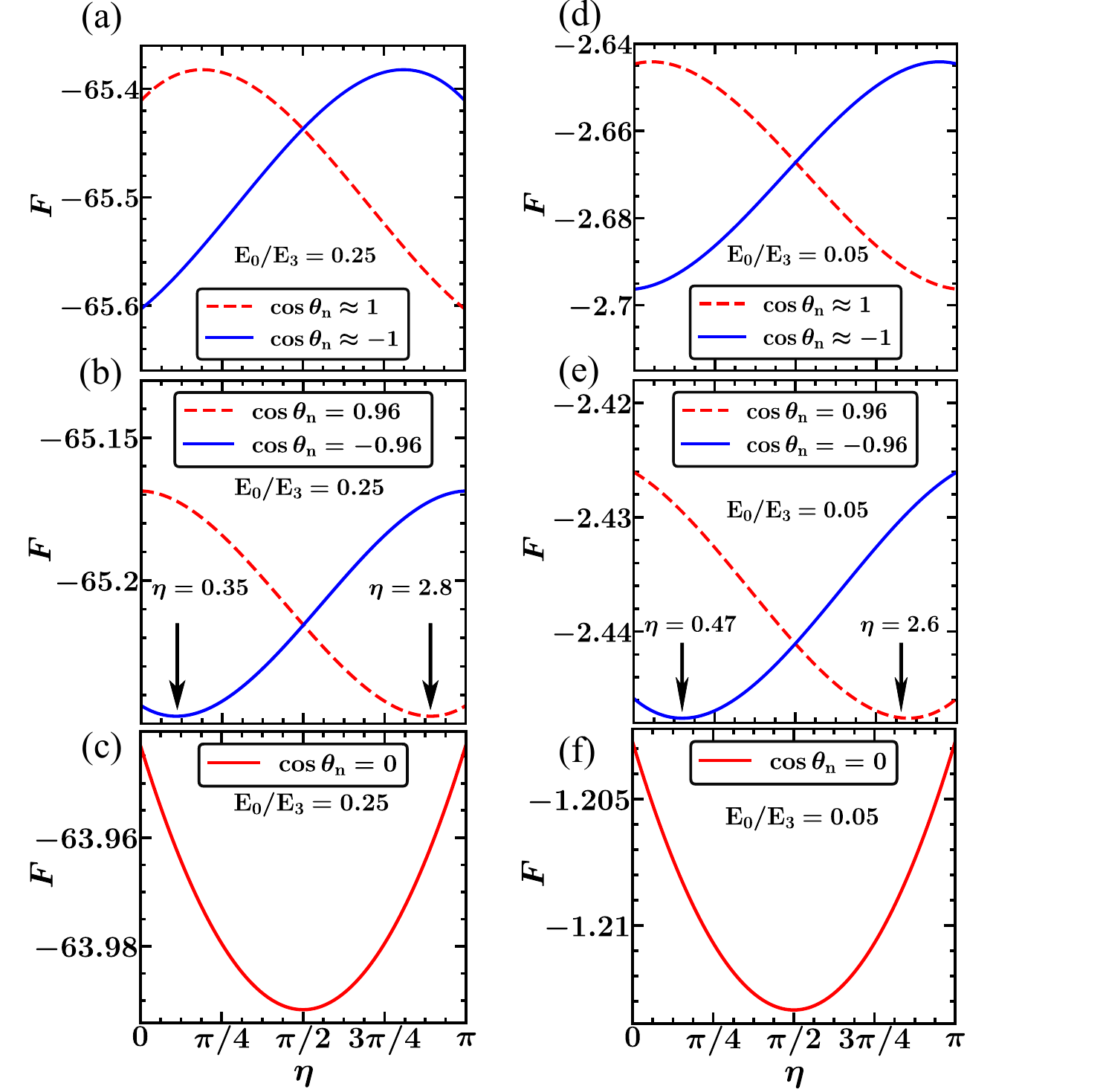}
		\caption{Free energy in the unit of $g_{11}$ {\it versus} $\eta$ for  (a--c) $E_0/E_3 = 0.25$ and (d--f) $E_0/E_3 = 0.05$ with  $\cos\theta_n = \pm 1$ (a and d), $\pm 0.96$ (b and e), and $0 $ (c and f). }
		\label{fig:plot1}
	\end{figure}

	Since Eqs.~\eqref{u_n} and \eqref{Theta} are coupled, they need to be solved simultaneously with the boundary conditions $\tilde{u}_n(0) =1$, $\tilde{u}_n(\infty)=0$, $\Theta (0) =\pi$, and $\Theta(\infty) =0$ for a possible skyrmion solution with polarization pointing down at its center and pointing up at very large distances. 
	By solving $\tilde{u}_n$ and $\Theta$ for chosen $\theta_n$ and $E_3$, we calculate $F$ (in the unit of $g_{11}$) at different values of $\eta$ for $c_{12}/c_{11} = 1/2$ because this ratio is approximately $0.45$ in PbTiO$_3$ systems \cite{Yuan2023}. We do not need to consider the values of any other parameters of the free energy because it can be explicitly expressed by the dimensionless parameter $E_0/E_3$ only.  Figure~\ref{fig:plot1} shows the variation of $F$ with $\eta$ at different values of $\theta_n$ and $E_3$. The free energy minimizes at different values of $\eta$ in the range $0 \leq \eta \leq \pi$ for $0 \leq \theta_n \leq \pi$; in particular, $\eta = \pi$ for $\cos(\theta_n) \approx 1$; $\eta = \pi/2$ for $\cos(\theta_n) =0$; $\eta = 0$ for $\cos(\theta_n) \approx -1$. Therefore, it is expected that at the top (bottom) layer, the skyrmion will be of Neel type with an inward (outward) direction of polarization. In the central layer, the skyrmion will proliferate as Bloch type. As $E_0/E_3$ decreases, the free energy increases, and it is found that for $E_0/E_3 \lesssim 1/22$, free energy becomes positive. Note that the free energy here is determined with respect to the global ferroelectric ground state energy, and skyrmion appears as a defect on this ground state only when the free energy becomes negative. Therefore, although the Eqs.~\eqref{u_n} and \eqref{Theta} provide a skyrmion solution, a skyrmion will not be proliferated for $E_3 \geq  22E_0$. 
	
	\begin{figure}
		\centering
		\includegraphics[width=0.97\linewidth]{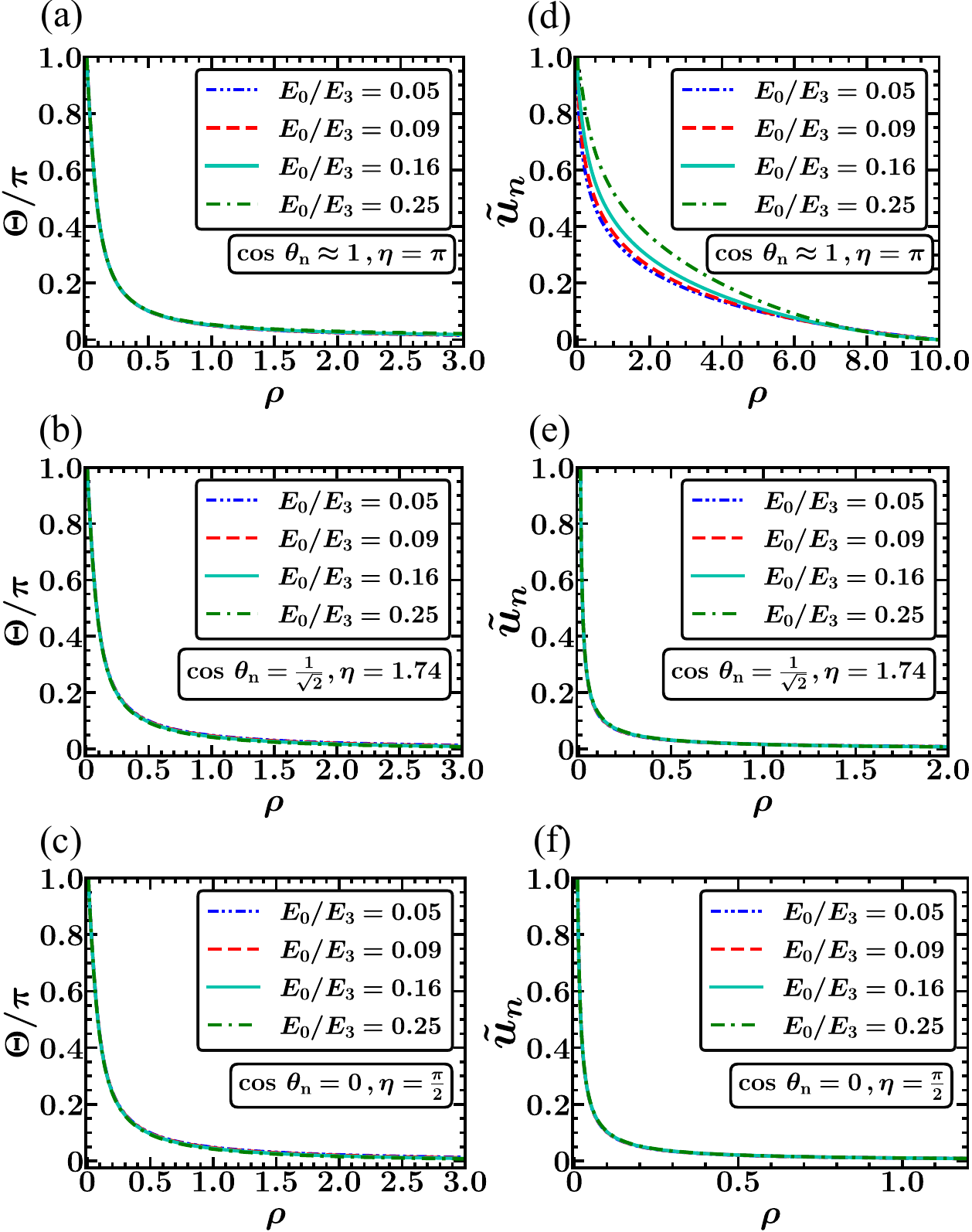}
		\caption{The solutions of $\Theta$ (a--c)  and $\tilde{u}_n$ (d--f) {\it versus} $\rho = r/r_0$ for different values of $E_0/E_3$, $\cos\theta_n$ and corresponding $\eta$ for which free energy is minimum (Fig.~1). The solutions of $\Theta$ are almost independent of $E_0/E_3$ and $\cos\theta_n$ as well. This proves $r_0$ as the most appropriate length scale as far as the skyrmions are concerned. A reasonably well-fitted form of the profile of $\Theta$ is found to be $\Theta \approx 4\arctan [\exp(-3.4\sqrt{\rho})]$.}
		\label{fig:plot2}
	\end{figure}
	
\begin{figure}
		\centering
		\includegraphics[width=\linewidth]{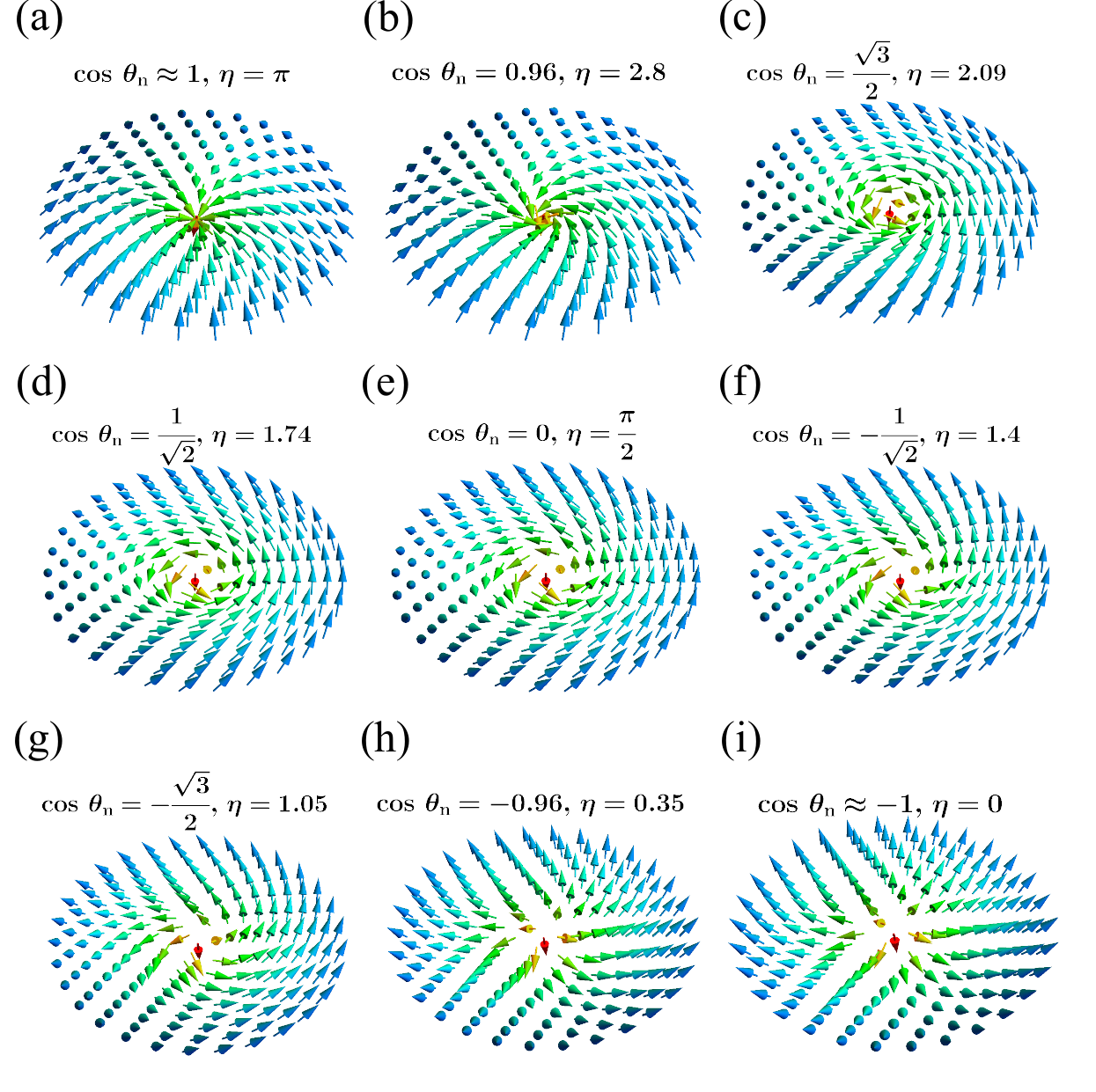}
		\caption{Structures of skyrmions in different layers aliasing with corresponding values of $\cos\theta_n$ and  $\eta$ for which free energies are the least. Structures continuously change from exclusively inward Neel type (a) to exclusively Bloch type (e) and to outward Neel type (i) as $\cos\theta_n$ changes  from $1\to 0\to -1$. }
		\label{fig:skdiagram}
	\end{figure}

	The solutions of $\tilde{u}_n(\rho)$ and $\Theta (\rho)$ in Eqs.~\eqref{u_n} and \eqref{Theta} for $\cos(\theta_n) \approx 1$ and $0$, suitable for top and central layers respectively, at different values of $E_0/E_3$ are shown in Fig.~\ref{fig:plot2}. The profile of $\Theta (\rho)$ is found to be independent of $E_0/E_3$, suggesting $r_0$ as the appropriate length scale for the skyrmion solutions. Therefore, as electric field increases, the radius of skyrmion shrinks. Using the profiles of $\Theta$ and the values of $\eta$ corresponding to $\cos(\theta_n) = 1$ and $0$, the structures of the skyrmions are obtained as Neel and Bloch types (Fig.~\ref{fig:skdiagram}) respectively. For the bottommost layer, $\cos(\theta_n) \approx -1$ and $\eta = 0$, the solutions of $\tilde{u}_n$ and $ \Theta $  are exactly the same as in Fig.~\ref{fig:plot2}(a). Therefore, the structure of the skyrmion in this case will also be Neel type, but the orientation of polarization vectors will be outward, opposite to the case of topmost layer. The skyrmion solutions are also obtained for other planes with $\cos\theta_n \neq 0,\pm 1$ where the structures (Fig.~\ref{fig:skdiagram}) interpolate  between Neel and Bloch types. However, the solutions with monotonic $\Theta$ for all values of $\cos\theta_n$ is found for $E_0/E_3 \lesssim 0.3$. Therefore skyrmion solutions are possible for the electric field $E_3 > 3.3 E_0$.

	We determine topological charge \cite{suppli} 
	\begin{equation}
		Q=  \frac{1}{4\pi} \int_0^R d\rho \int_0^{2\pi} 
		d\phi\, \sin\Theta \, \left( \frac{d\Theta}{d\rho}\right)\left( \frac{d\Phi}{d\phi}\right)
	\end{equation}
	and find that $Q\approx -1$ for the skyrmions in each layer with $R=10$; $Q=-1$ for $R\to \infty$. Readers should not be confused with negative $Q$ for skyrmions; it is negative because the polarization direction at the center of the skyrmion is downward and it is upward at large distances. The difference between a skyrmion and an antiskyrmion is not due to the sign of $Q$. These are differentiated by how azimuthal angle, $\Phi$, of the polarization vector depends on planar polar angle $\phi$. The positive (negative) sign of $d\Phi /d\phi$ indicates a skyrmion (an antiskyrmion).

	As found above, the skyrmions proliferate for the range of Electric field $3.3 E_0 < E_3 < 22E_0$.  The lower bound here is underestimated because another competing state, {\it i.e.}, possible spiral phase may occur at low electric field regime. Similarly, the upper bound is possibly overestimated because some other phase may occur between a phase of isolated skyrmions and the ferroelectric phase. Nonetheless, the upper bound here is about seven times that of the lower bound, in consistence with recent simulation where it is found to be about five times \cite{Yuan2023} at zero temperature. 
	By considering the parameters \cite{Yuan2023,WANG20052495,LI2002} relevant for PTO, $c_{11} = 1.7 \times 10^{11}$ N/m$^2$, $g_{11} = 2.8 \times 10^{-10}$ Nm$^4$/C$^2$, $q_{44} = 7.5 \times 10^{9}$ Nm$^2$/C$^2$, and polarization $P_0 = 0.3$ C/m$^2$, we find $U_0 = \sqrt{g_{11}/c_{11}}P_0 = 1.2\times 10^{-11}$ m,  and $E_0 = U_0^2 q_{44}^2 P_0/g_{11} = 8.7 \times 10^6$ V/m, by reinstating the magnitude of polarization $P_0$. Therefore, the lower bound of electric field for observing skyrmions would be $E_3 =2.9 \times 10^7$ V/m which has been underestimated for the above-mentioned reason and the inclusion of depolarizing field as well. Nonetheless, this estimate agrees quite well with the estimated value of electric field in numerical simulation \cite{HONG2018155}. For this lower bound of $E_3$, the length scale $r_0 = U_0q_{44}P_0/E_3$ is found to be $1.3$ nm, suggesting the diameter of a skyrmion is in nano-meter scale.

	The fact that skyrmions are only possible for a range of electric field is corroborated with the experimental observation of skyrmions in PTO layers in (PTO)$_n$/(STO)$_n$ heterostructures for  $12 \leq n \leq 18$; because for lesser (greater) $n$, the depolarizing field is more (less) causing net electric field beyond the range for which skyrmions may be observed \cite{Hong2017}. Since $\cos\theta_n = 1$ (within 1\%) for $\theta_n < 8^o$, the approximation of $\cos\theta_n \approx 1$ for the topmost layer is partially justified. However, as the free energy remains minimum approximately at $\eta = \pi$ for $\cos\theta_n > 0.97$, the skyrmion at the topmost layer will still be Neel type with inward orientation.

	Apart from the intrinsic strain in the lattice, extrinsic strain arises due to the application of electric field in the PTO plane. Since the applied electric field, together with the depolarizing field, is responsible for electric polarization, the strain couples with polarization giving rise to the interaction term in free energy $\mathcal{F}_{\rm int}$. However, as strain is a symmetric tensor, it fails to provide antisymmetric interaction similar to  DMI \cite{DMI,Moriya1960}. Nonetheless, as we have found, the necessary gradient of polarization in the interaction term sets the stage for proliferating topological structures like skyrmions. Recently, {\it ab initio} calculations \cite{Zhao2021} have made prediction of realizing DMI in certain ferroelectric systems. 
	We anticipate that the tensile strain which is an antisymmetric tensor would be responsible for generating such an interaction. 

	We now discuss about the possibility of antiskyrmion structure. We find that $F$ possesses degenerate minima \cite{suppli} at $\eta = 0$ and $\pi$ for all $\theta_n$ , implying the equal possibility of inward and outward Neel type antiskyrmion structures. Therefore, the configurations in favor of antiskyrmions are mutually get canceled due to the lack of any symmetry breaking effect which could allow only one of these two possibilities. This rules out the possibility of proliferating antiskyrmion in any of the layers.

	We have identified the cause of forming polar skyrmion structures in PTO layers in the absence of any {\it a priori} known interaction involving electric polarizations with linear gradients so as to cause orientation at different directions at different spatial points. The accomplished polar skyrmion solution linked with the elastic displacement vector for defects in PTO layers and its structural variations from layer to layer have been found to be consistent with the experiments. The range of electric field for which skyrmions can be realized is estimated in reasonable agreement with previously reported numerical simulation.
		

%

	\clearpage
	
	\setcounter{equation}{0}
	\renewcommand{\theequation}{SE\arabic{equation}}
	\setcounter{figure}{0}
	\renewcommand{\thefigure}{SF\arabic{figure}}
	\setcounter{section}{0}
	\renewcommand{\thesection}{S\arabic{section}}  

	\begin{widetext}
		\centering
		\Large{\bf{ Supplementary Material for ``Theory of Polar Skyrmions in Layered Structure of Ferroelectric Perovskites"} }\\[1cm]
	\end{widetext}
	
	\noindent This Supplemental Material consists of four sections:
	\ref{Free_Energy}, The free energy is derived in the form of polar variables, 
	\ref{sk_euler}, Derivation of the Euler equations pertaining to a skyrmion solution, 
	\ref{Anti-Sk}, Euler equations pertaining to antiskyrmion and their solutions, and 
	\ref{charge}, Expression of topological charge in terms of polar variables.
	
	\section{Free Energy}\label{Free_Energy}  
	In this section, we explicitly transform the different contributions to free energy from Cartesian to polar system with $x = r \cos\phi$, $y=r \sin\phi$, $P_1 = \sin\Theta (r) \cos \Phi (\phi)$, $  P_2 = \sin\Theta (r) \sin \Phi (\phi)$, and $P_3 = \cos\Theta (r)$. In polar form, the elastic displacement vector $\bm{u} = u_n(r) \left[\sin\theta_n\cos\phi,\, \sin\theta_n \sin\phi,\, \cos\theta_n \right]$ (see Eq.(2)).
	
	The spatial derivatives of the $P_i$'s and $u_i$'s in the polar form are found as    
	\begin{eqnarray}
		&&	\frac{\partial P_1}{\partial x_1} = \cos\phi\frac{\partial}{\partial r}\left[\sin\Theta\cos\Phi\right]-\frac{\sin\phi}{r}\frac{\partial}{\partial \phi}\left[\sin\Theta\cos\Phi\right] \nonumber \\
		&&\hspace{0.3cm} = \cos \phi \cos \Phi \cos \Theta \Theta_r+\frac{1}{r}\sin\phi\sin\Theta\sin\Phi\Phi_{\phi}\label{delp1delx1} \, ,
	\end{eqnarray}
	\begin{eqnarray}
		&&	\frac{\partial P_2}{\partial x_2} = \sin\phi\frac{\partial}{\partial r}\left[\sin\Theta\sin\Phi\right]+\frac{\cos\phi}{r}\frac{\partial}{\partial \phi}\left[\sin\Theta\sin\Phi\right]\nonumber \\
		&& \hspace{0.3cm} = \sin\phi\cos\Theta\sin\Phi\Theta_r+\frac{1}{r}\cos\phi\sin\Theta\cos\Phi\Phi_{\phi}\label{delp2delx2}\, ,
	\end{eqnarray}
	\begin{eqnarray}
		&&	\frac{\partial P_1}{\partial x_2} = \sin\phi\frac{\partial}{\partial r}\left[\sin\Theta\cos\Phi\right]+\frac{\cos\phi}{r}\frac{\partial}{\partial \phi}\left[\sin\Theta\cos\Phi\right] \nonumber \\
		&& \hspace{0.3cm} = \sin\phi\cos\Theta\cos\Phi\Theta_r-\frac{1}{r}\cos\phi\sin\Theta\sin\Phi\Phi_{\phi}\label{delp1delx2} \, ,
	\end{eqnarray}
	\begin{eqnarray}
		&&	\frac{\partial P_2}{\partial x_1} = \cos\phi\frac{\partial}{\partial r}\left[\sin\Theta\sin\Phi\right]-\frac{\sin\phi}{r}\frac{\partial}{\partial \phi}\left[\sin\Theta\sin\Phi\right]\nonumber \\
		&& \hspace{0.3cm} = \cos\phi\cos\Theta\sin\Phi\Theta_r-\frac{1}{r}\sin\phi\sin\Theta\cos\Phi\Phi_{\phi}\label{delp2delx1}\,, 
	\end{eqnarray}
	\begin{eqnarray}
		&& \hspace{-1.6cm}	\frac{\partial P_3}{\partial x_1} = \cos\phi\frac{\partial}{\partial r}\left[\cos\Theta\right]-\frac{\sin\phi}{r}\frac{\partial}{\partial \phi}\left[\cos\Theta\right] \nonumber \\
		&&  \hspace{-0.8cm} =-\cos\phi\sin\Theta\Theta_r\label{delp3delx1}\,,
	\end{eqnarray}
	\begin{eqnarray}
		&& \hspace{-1.6cm}	\frac{\partial P_3}{\partial x_2} = \sin\phi\frac{\partial}{\partial r}\left[\cos\Theta\right]+\frac{\cos\phi}{r}\frac{\partial}{\partial \phi}\left[\cos\Theta\right] \nonumber \\
		&& \hspace{-0.8cm} = -\sin\phi\sin\Theta\Theta_r\label{delp3delx2}\,,
	\end{eqnarray}    
	\begin{eqnarray}
		&&	\frac{\partial u_1}{\partial x_1} = \cos\phi  \frac{\partial}{\partial r} \left[u_n \cos\phi \sin\theta _n\right]-
		\frac{1}{r}\sin\phi  \frac{\partial}{\partial \phi } \left[u_n \cos\phi  \sin\theta _n\right]\nonumber \\ 
		&& \hspace{0.5cm} = \cos ^2\phi \sin\theta _n \frac{\partial u_n}{\partial r}+\frac{1}{r}u_n \sin ^2\phi \sin \theta _n\label{delu1delx1}\,,
	\end{eqnarray}
	\begin{eqnarray}
		&&	\frac{\partial u_2}{\partial x_2} = \sin\phi  \frac{\partial}{\partial r}\left[u_n \sin \phi  \sin \theta _n\right]
		+\frac{1}{r}\cos\phi  \frac{\partial}{\partial \phi } \left[u_n \sin\phi \sin \theta _n\right] \nonumber\\
		&& \hspace{0.5cm} = \sin ^2\phi \sin\theta _n\frac{\partial u_n}{\partial r}+\frac{1}{r}u_n \cos ^2\phi \sin\theta _n\label{delu2delx2}\,,
	\end{eqnarray}
	\begin{eqnarray}
		&&	\frac{\partial u_1}{\partial x_2} = \sin\phi  \frac{\partial}{\partial r} \left[u_n \cos\phi  \sin\theta _n\right]  
		+  \frac{1}{r}\cos\phi \frac{\partial}{\partial \phi } \left[u_n \cos\phi  \sin\theta _n\right] \nonumber \\ 
		&& \hspace{0.5cm} = \sin\phi\cos\phi\sin\theta_{n}\frac{\partial u_n}{\partial r}-\frac{1}{r}\sin\phi\cos\phi\sin\theta_n u_n\label{delu1delx2}\,,
	\end{eqnarray}
	\begin{eqnarray}
		&&	\frac{\partial u_2}{\partial x_1} = \cos\phi\frac{\partial}{\partial r} \left[u_n \sin\phi  \sin\theta _n\right]  
		- \frac{1}{r}\sin\phi  \frac{\partial}{\partial \phi } \left[u_n \sin\phi  \sin\theta _n\right] \nonumber \\
		&&\hspace{0.5cm} = \cos\phi\sin\phi\sin\theta_{n}\frac{\partial u_n}{\partial r}-\frac{1}{r}\sin\phi\cos\phi\sin\theta_n u_n\label{delu2delx1}\,,
	\end{eqnarray}
	\begin{eqnarray}
		&&	\hspace{-1.0cm}\frac{\partial u_3}{\partial x_1} = \cos\phi  \frac{\partial}{\partial r} \left[u_n \cos\theta _n\right]- \frac{1}{r}\sin\phi \frac{\partial}{\partial \phi } \left[u_n \cos\theta _n\right] \nonumber \\
		&&\hspace{-0.2cm} = \cos\phi  \cos\theta _n \frac{\partial u_n}{\partial r}\label{delu3delx1}\,,
	\end{eqnarray}
	\begin{eqnarray}
		&&\hspace{-1.0cm} 	\frac{\partial u_3}{\partial x_2} = \sin\phi \frac{\partial}{\partial r} \left[u_n \cos\theta _n\right]+\frac{1}{r}\cos\phi \frac{\partial}{\partial \phi } \left[u_n \cos\theta _n\right] \nonumber \\
		&&\hspace{-0.3cm} =  \sin\phi \cos\theta _n \frac{\partial u_n}{\partial r}\label{delu3delx2} \, . 
	\end{eqnarray}
	
	We now formulate different components of free energy in terms of polar variables using above derivatives.\\
	
	\noindent{\bf Gradient energy:}
	The gradient free energy density is given by
	\begin{eqnarray}
		\mathcal{F}_{\rm gr} &=& \frac{1}{2}g_{ijkl}P_{i,j}P_{k,l} \, .\nonumber
	\end{eqnarray}
	Using Eqs.~\eqref{delp1delx1}--\eqref{delp3delx2}, and symmetry properties of $g_{ijkl}$ we find,
	\begin{eqnarray}
		&&\mathcal{F}_{\rm gr}= \frac{g_{11}}{4r^{2}}\biggl[\left(\frac{\partial \Phi}{\partial \phi}\right)^{2}+2\left(\frac{\partial \Theta}{\partial \phi}\right)^{2}+r^{2}\left[\left(\frac{\partial \Phi}{\partial r}\right)^{2}  + 2\left(\frac{\partial \Theta}{\partial r}\right)^{2}    \right] \nonumber \\
		&&  \hspace{0.8cm} -\biggl[\left(\frac{\partial \Phi}{\partial \phi}\right)^{2}+r^{2}\left(\frac{\partial \Phi}{\partial r}\right)^{2}\biggr]\cos 2\Theta\biggr] \, .  \label{f_gr1}
	\end{eqnarray} 
	As $\frac{\partial \Theta (r)}{\partial \phi} = 0$ and $\frac{\partial \Phi (\phi)}{\partial r} = 0$, Eq.~\eqref{f_gr1} finally reduces to 
	\begin{equation}
		\mathcal{F}_{\rm gr} = \frac{g_{11}}{2}(\Theta_{r}^{2}+\frac{\sin^{2}\Theta}{r^{2}}\Phi_\phi^2) \, . \label{f_gr}
	\end{equation}
	
	\noindent{\bf Strain Energy :}
	Different components of strain tensors are as follows.
	Using Eqs.~\eqref{delu1delx1}--\eqref{delu3delx2}, we find
	\begin{eqnarray}
		&&\hspace{-2.3cm}\epsilon _{11} = \frac{\partial u_1}{\partial x_1} \nonumber \\
		& & \hspace{-1.8cm} =\cos ^2\phi \sin\theta _n \frac{\partial u_n}{\partial r}+\frac{1}{r}u_n \sin ^2\phi \sin \theta _n\,,
	\end{eqnarray}
	\begin{eqnarray}
		&&\hspace{-2.3cm} \epsilon _{22} = \frac{\partial u_2}{\partial x_2} \nonumber \\
		& & \hspace{-1.8cm} = \sin ^2\phi \sin\theta _n\frac{\partial u_n}{\partial r}+\frac{1}{r}u_n \cos ^2\phi \sin\theta _n\,,
	\end{eqnarray}
	\begin{eqnarray}
		&&\hspace{-1.5cm}\epsilon _{12} = \frac{1}{2}\left(\frac{\partial u_1}{\partial x_2}+\frac{\partial u_2}{\partial x_1}\right) \nonumber \\   
		&&\hspace{-1.0cm} = \sin\phi  \cos\phi  \sin\theta _n \frac{d u_n}{d r}-\frac{1}{r}u_n \sin\phi  \cos\phi  \sin\theta _n\,,
	\end{eqnarray}
	\begin{eqnarray} 
		&&\hspace{-5.5cm}\epsilon _{23} = \frac{1}{2}\left(\frac{\partial u_3}{\partial x_2}+\frac{\partial u_2}{\partial x_3}\right) \nonumber \\
		&&\hspace{-5.0cm} =\frac{1}{2} \sin\phi \cos\theta _n \frac{\partial u_n}{\partial r}\,,
	\end{eqnarray}
	\begin{eqnarray}
		&&\hspace{-5.5cm}\epsilon _{13} = \frac{1}{2}\left(\frac{\partial u_3}{\partial x_1}+\frac{\partial u_1}{\partial x_3}\right) \nonumber \\  
		&&\hspace{-5.0cm}=\frac{1}{2} \cos\phi  \cos\theta _n \frac{\partial u_n}{\partial r}\,,
	\end{eqnarray}
	with $\partial u_1/\partial x_3 = \partial u_2/\partial x_3 = 0$ and $\epsilon_{33}=0$ for a two-dimensional layer. 
	Therefore, the elastic energy is found to be
	\begin{eqnarray}
		\mathcal{F}_{\rm st} &=& \frac{1}{2} c_{11} \left(\epsilon _{11}^2+\epsilon _{22}^2\right)+c_{12} \epsilon _{11} \epsilon _{22} 
		+ 2c_{44} \left(\epsilon _{12}^2+\epsilon _{13}^2+\epsilon _{23}^2\right) \nonumber \\
		&=& \sin^{2}\theta_{n} \left[ c_{11} \frac{u_{n}^{2}}{2r^{2}}+c_{12}\frac{u_{n} }{r}\left( \frac{\partial u_n}{\partial r}\right)\right] \nonumber \\
		&&-\frac{1}{4}\left[ \cos^{2}\theta_{n} (c_{11}+c_{12}) -2c_{11} \right] \left( \frac{ \partial u_{n}}{\partial r} \right)^2 \, .\label{st_energy}
	\end{eqnarray}
	\noindent{\bf Interaction Energy :}
	We here show the derivation of the $\mathcal{F}_{\rm int}$ which is given in Eq.~(4) of the main text. Expanded Eq.~(1) of the main text in the Cartesian coordinate, we find  
	\begin{eqnarray}
		\mathcal{F}_{\rm int} 
		&=& q_{44} \left[ u_1 P_1 \frac{\partial P_2}{\partial x_2}+u_1 P_2 \frac{\partial P_1}{\partial x_2}+ u_2 P_1 \frac{\partial P_2}{\partial x_1}+ u_2 P_2 \frac{\partial P_1}{\partial x_1} \right. \nonumber \\
		&& + \left. u_3 P_2 \frac{\partial P_3}{\partial x_2}+ u_3 P_3 \frac{\partial P_2}{\partial x_2}+ u_3 P_1 \frac{\partial P_3}{\partial x_1}+ u_3 P_3 \frac{\partial P_1}{\partial x_1} \right]\,.\nonumber 
	\end{eqnarray}
	Using Eqs.~\eqref{delp1delx1}--\eqref{delp3delx2} and further simplifying, we obtain
	\begin{eqnarray}  
		\mathcal{F}_{\rm int}&=&\frac{q_{44}u_n }{4}\biggl[\sin\theta_n\cos [2(\Phi-\phi)] 
		\biggl(\sin 2\Theta\Theta_{r}+\dfrac{2\sin^{2}\Theta \Phi_\phi}{r} \biggr)\nonumber \\
		&& + 4\cos\theta_n \cos(\Phi-\phi)\biggl(\cos 2\Theta\,\Theta_{r}
		+ \dfrac{\sin 2\Theta \Phi_\phi}{2r} \biggr)  \nonumber \\ 
		&&-\sin\theta_n\cos\left[2(\Phi+\phi)\right]\biggl( \sin 2\Theta\Theta_r 
		- \dfrac{2\sin^{2}\Theta \Phi_\phi}{r}\biggr)     
		\biggr] \, .\nonumber \\
		\label{Fener_int} 
	\end{eqnarray}
	where $\Theta_r = \frac{d\Theta}{dr}$ and $\Phi_\phi = \frac{d\Phi}{d\phi}$.
	
	\noindent{\bf Electrical Energy:}
	The free energy due to the application of electric field is given by
	\begin{equation}
		\mathcal{F}_{\rm elec} = -\bm{E}\cdot \bm{P} = -E_3 \cos\Theta  \, . \label{free_elec}
	\end{equation}

	\begin{widetext}
		\section{EULER EQUATION FOR THE SKYRMION SOLUTION}\label{sk_euler}
		The total free energy  of the system is then given by
		\begin{equation}
			F =  \int dr \int d\phi\,\, r \left[ \mathcal{F}_{\rm gr} + \mathcal{F}_{\rm st} + \mathcal{F}_{\rm int}+ \mathcal{F}_{\rm elec} \right] 
		\end{equation} 
		where the different free energy densities are expressed in Eqs.~\eqref{f_gr}, \eqref{st_energy}, \eqref{Fener_int} and \eqref{free_elec}. We have ignored the last term in Eq.~\eqref{Fener_int}, because only the skyrmion solution is considered here. 
		
		The derivatives of $\mathcal{F}$ with respect to $\Theta$, $\Theta_r$, $u_n$ and $\frac{du_n}{dr}$ are found to be
		\begin{eqnarray}
			\hspace{-3cm} \frac{\partial F}{\partial \Theta} &=& \int dr \int d\phi \,  r \biggl[E_3 \sin\Theta+\frac{g_{11} \sin\Theta \cos\Theta }{r^2}
			+ \frac{1}{2} q_{44} \cos 2\eta\,u_n \sin\theta _n \biggl(-\Theta_r \sin ^2\Theta+\Theta_r \cos ^2\Theta\nonumber\\
			&& +\frac{2 \sin\Theta\cos\Theta}{r}\biggr)+q_{44}\cos\eta\,u_n\cos\theta _n \biggl(-2 \Theta_r \sin 2\Theta
			- \frac{\sin ^2\Theta}{r}+\frac{\cos ^2\Theta}{r}\biggr)\biggr]\label{delFdeltheta}\,.
		\end{eqnarray}
		\begin{eqnarray}
			\hspace{-1.5cm}\frac{\partial F}{\partial \Theta_r} =\int dr\int d\phi \, r \biggl[g_{11} \Theta_r+q_{44} \cos\eta  u_n \cos\theta _n \cos2 \Theta+ \frac{1}{2} q_{44} \cos 2\eta 
			\,u_n \sin\theta _n\sin\Theta\cos\Theta\biggr] \label{delFdelthetar}\,.
		\end{eqnarray}
		\begin{eqnarray}
			\hspace{-2cm} \frac{\partial F}{\partial u_n} &=& \int dr \int d\phi \, r \biggl[\frac{c_{11}  \sin ^2\theta _n}{r^2}u_n+\frac{c_{12} \sin ^2\theta _n }{r}\frac{du_n}{dr}+\frac{1}{2} q_{44} \cos 2\eta
			\sin\theta _n \biggl(\Theta_r \sin\Theta\cos\Theta+\frac{\sin ^2\Theta}{r}\biggr) \nonumber \\
			&&+q_{44} \cos\eta \cos\theta _n
			\biggl(\Theta_r \cos 2 \Theta+\frac{\sin\Theta \cos\Theta}{r}\biggr)\biggr]\,.\label{delFdelun}
		\end{eqnarray}
		\begin{eqnarray}
			\hspace{-4cm}	\frac{\partial F}{\partial \left(du_n/dr\right)} = \int dr\int d\phi \, r \biggl[\frac{c_{12}  \sin ^2\theta _n}{r}u_n-\frac{1}{4} \frac{d u_n}{d r} \biggl[2 c_{12} \cos ^2\theta _n+c_{11}
			\bigl(\cos 2 \theta _n- 3\bigr)\biggr]\biggr]\,.\label{delFdelunprime}
		\end{eqnarray}
		Using Eqs. ~\eqref{delFdeltheta}--\eqref{delFdelunprime}, we find the Euler equations for $u_n$ and $\Theta$ respectively as
		\begin{eqnarray}
			&&\hspace{-1.5cm}	\biggl[\cos^{2}\theta_n c_{12}+c_{11}\bigl(\cos^{2}\theta_n-2\bigr)\biggr]\biggl(\frac{d^{2}u_{n}}{dr^{2}}+\frac{1}{r}\frac{du_{n}}{dr}\biggr)
			+ \frac{2c_{11}}{r^{2}}\sin^{2}\theta_n u_{n}  \nonumber \\
			&& \hspace{-0.9cm} =-q_{44}\biggl[\bigl(2\cos\eta\cos\theta_n\cos 2\Theta
			+\frac{1}{2}\cos 2\eta\sin\theta_n \sin 2\Theta\bigr)\Theta_r   +\frac{1}{r}\bigl(\cos 2\eta\sin\theta_n\sin^{2}\Theta
			+ \cos\eta\cos\theta_n\sin 2\Theta\bigr)\biggr]\label{un_eq}\, ,
		\end{eqnarray}
		\begin{eqnarray}	
			g_{11}\biggl(\Theta_{rr}+\frac{\Theta_{r}}{r}-\frac{\sin 2\Theta}{2 r^2}\biggr) = E_3 \sin \Theta-\frac{q_{44}}{4}\biggl[\cos 2\eta \sin\theta_n 
			\sin 2\Theta \biggl(\frac{d u_n}{d r}-\frac{u_n}{r}\biggr)+4\cos\eta\cos\theta_n\cos 2\Theta\frac{d u_n}{d r}\biggr]\label{theta_eq}\,.
		\end{eqnarray}
		where $\Theta_{rr} = \frac{d^{2}\Theta}{dr^{2}}$ and $\Theta_r = \frac{d\Theta}{dr}$. Using dimensionless variables $\rho$ and electric field scale $E_0$ as described in the main text, we  convert Eqs.~\eqref{un_eq} and \eqref{theta_eq} in the forms of  Eqs.~(5) and (6) respectively.
	\end{widetext}
	
	\section{ANTISKYRMION SOLUTION AND ITS ENERGY}\label{Anti-Sk}
	
	As shown in the section \ref{sk_euler}, we have derived the corresponding Euler equation for obtaining the antiskyrmion solution here. In the expression of $F$, we have excluded the first two terms in Eq.~(4) as we are interested to find Euler equations for an antiskyrmion solution. The Euler equations for $\tilde{u}_n$ and $\Theta$  in terms of the dimensionless variable $\rho$ are found as 
	
	\begin{widetext}
		\begin{eqnarray}
			&& \hspace{-2cm}	\left[2-\cos^{2}\theta_{n}\left( 1+\frac{c_{12}}{c_{11}} \right) \right]\left(\dfrac{d^{2}\tilde{u}_n}{d\rho^{2}}+\dfrac{1}{\rho}\dfrac{d\tilde{u}_n}{d \rho}\right) 
			- 2\sin^{2}\theta_{n}\dfrac{\tilde{u}_n}{\rho^{2}} = -\dfrac{E_{0}^{\prime }}{E_{3}}\biggl[\sin \theta_{n} \cos (2\eta) 
			\left(\frac{\sin^{2}\Theta}{\rho}+\frac{\sin (2\Theta)}{2}\Theta_{\rho}\right)  \biggr]  \label{u_n} 
		\end{eqnarray}
		\begin{eqnarray}	 
			&&\hspace{-6.5cm}	\dfrac{d^{2}\Theta}{d \rho^{2}}+\dfrac{1}{\rho}\dfrac{d \Theta}{d \rho}-\dfrac{\sin 2\Theta}{2\rho^{2}} =-\dfrac{E_{0}}{E_{3}} 
			\biggl[\dfrac{1}{4} \cos 2\eta \sin 2 \Theta \sin \theta_{n}
			\biggl(\dfrac{\tilde{u}_n}{\rho}-\dfrac{ d \tilde{u}_n}{d \rho}\biggr) -\sin \Theta\biggr] \label{Theta} \, .
		\end{eqnarray}
	\end{widetext}
	
	\begin{figure}
		\centering   
		\includegraphics[width = \linewidth]{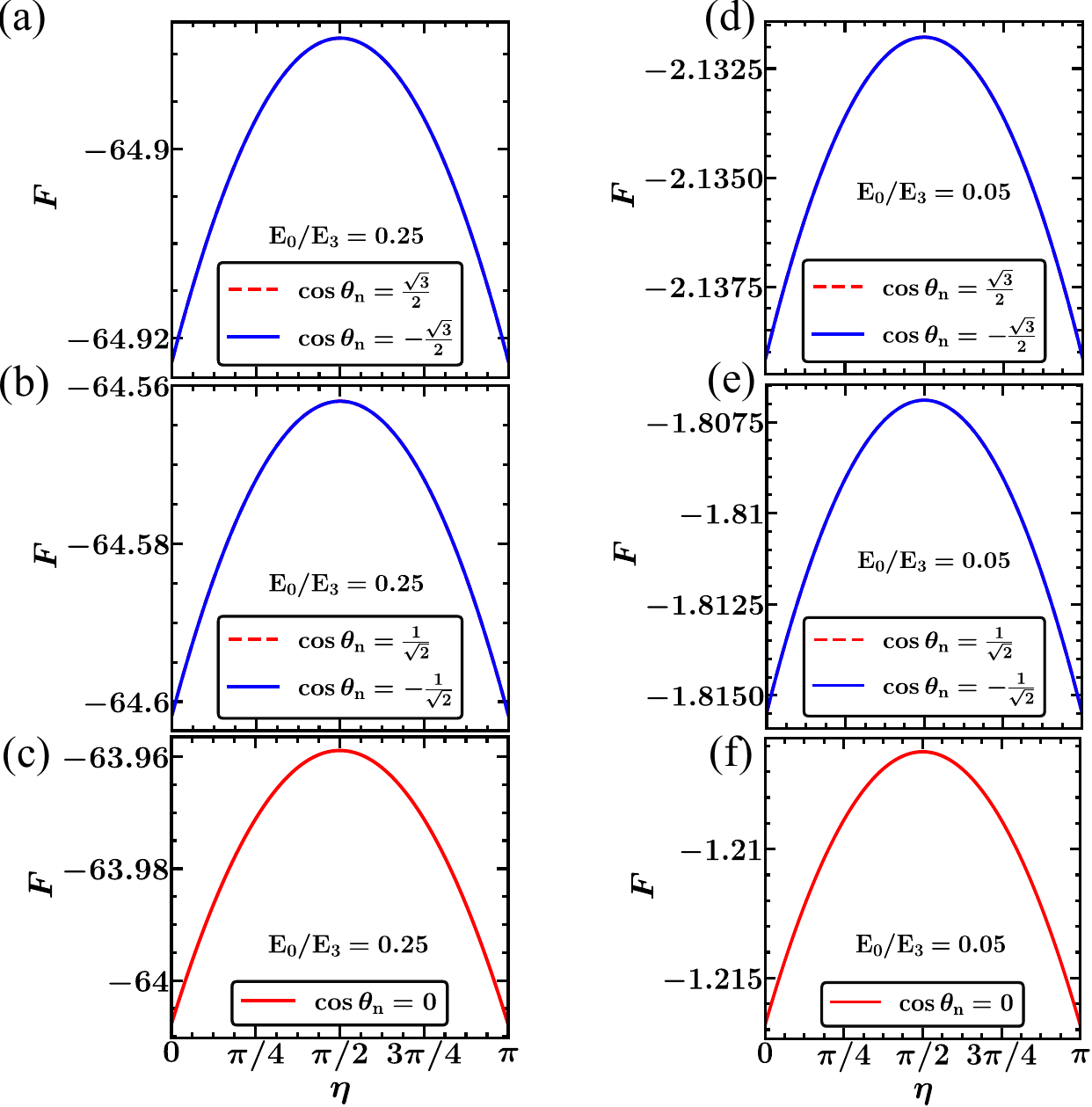}
		\caption{Free energy in the unit of $g_{11}$ \textit{versus} $\eta$ for (a-c) $E_0/E_3 = 0.25$ and (d-f) $E_0/E_3 = 0.05$ with $\cos\theta_{n} = \pm \frac{\sqrt{3}}{2}$(a and d), $\pm \frac{1}{\sqrt{2}}$(b and e), and 0 (c and f).} 
		\label{fig:plot3}
	\end{figure}
	\begin{figure}
		\centering    
		\includegraphics[width = \linewidth]{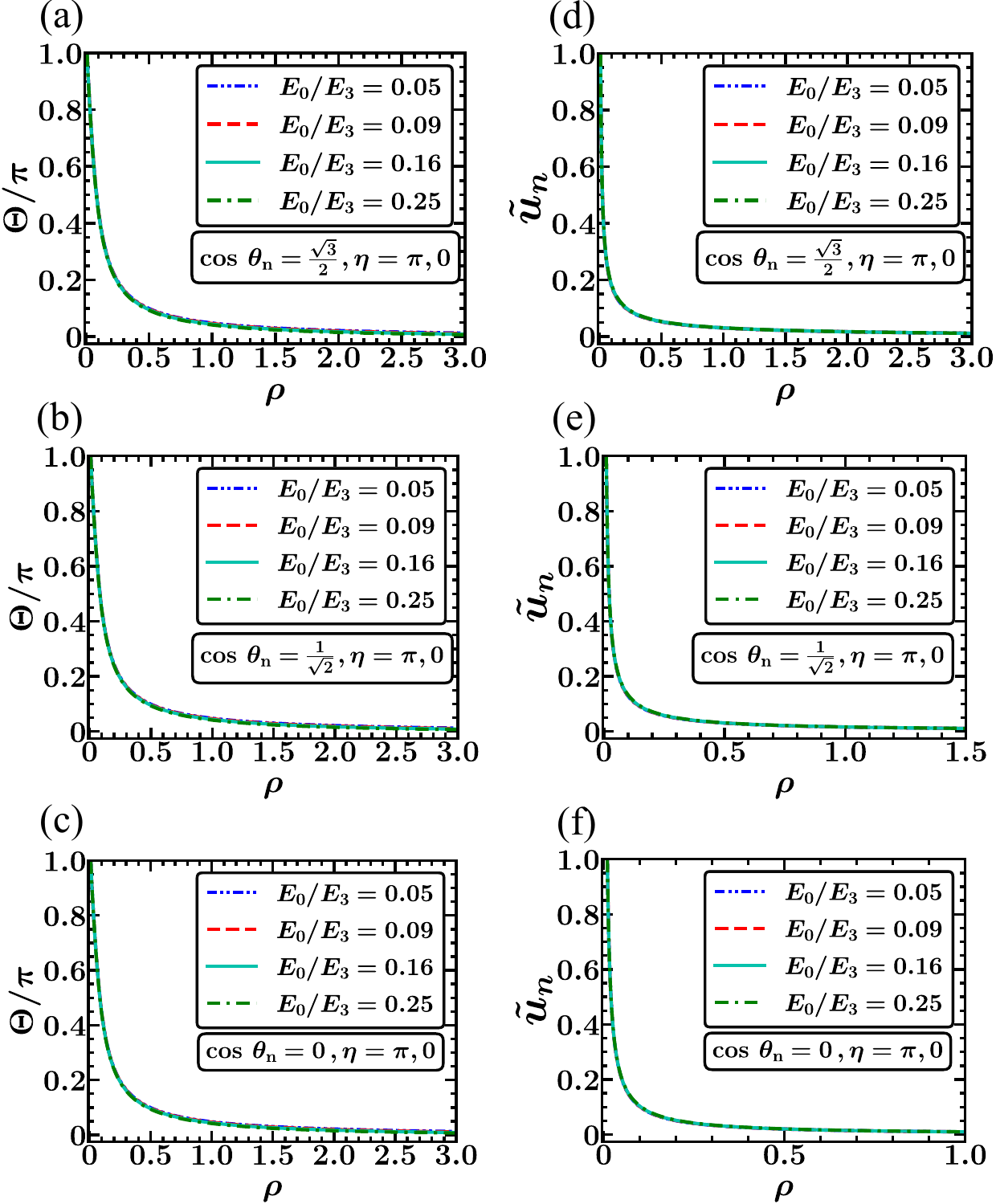}     
		\caption{The solutions of $\Theta$(a-c) and $\tilde{u}_{n}$(d-f) \textit{versus} $\rho = r/r_0$ for different values of $E_0/E_3$, $\cos\theta_{n}$ and corresponding $\eta$ for which free energy is minimum (a-f). The solutions of $\Theta$ are almost independent of $E_0/E_3$ and $\cos\theta_n$ as well.}     
		\label{fig:plot4}         
	\end{figure}
	\begin{figure}
		\centering   
		\includegraphics[width = \linewidth]{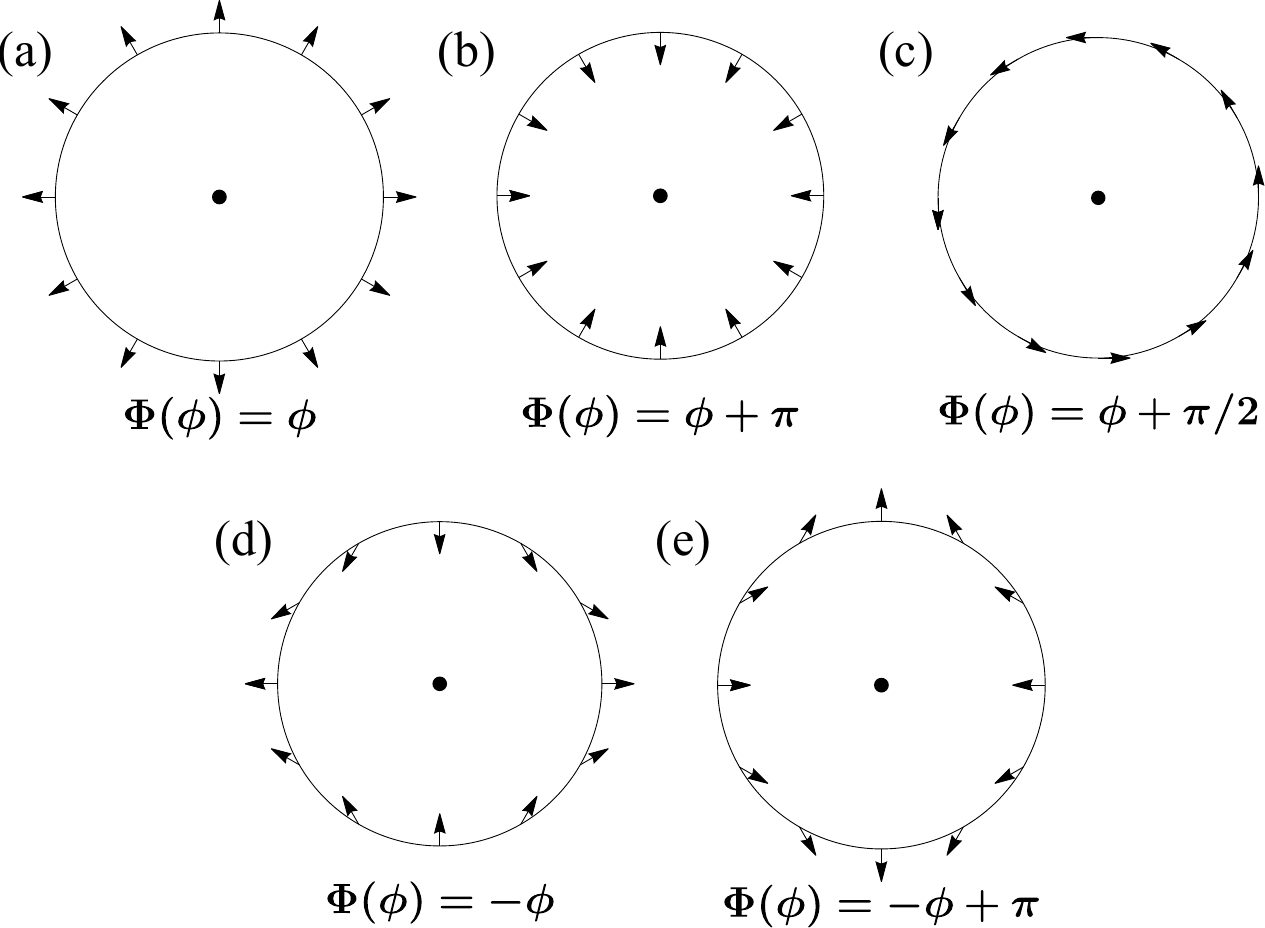}
		\caption{Direction of the polarization vector in a plane at a fixed radius $r$ for different functional form of $\Phi(\phi)$. $\Phi(\phi) = \phi$, $\Phi(\phi) = \phi+\pi$ and $\Phi = \phi+\frac{\pi}{2}$ corresponds to outward Neel type (a), inward Neel type(b) and Bloch type (c) skyrmions, respectively. (d) $\Phi = -\phi$ and (e) $\Phi = -\phi+\pi$ correspond to Neel type antiskyrmions with exactly opposite orientations of polarizations.}  \label{sk_antisk}
	\end{figure} 
	Eqs.~\eqref{u_n} and \eqref{Theta} have been solved simultaneously using boundary condition  condition $\tilde{u}_n(0) = 1$, $\tilde{u}_n(\infty) = 0$, $\Theta(0) = \pi$ and $\Theta(\infty) = 0$ for searching a possible antiskyrmion solution with polarization at its centre pointing downwards and pointing upwards at very large distances. Using the solutions of Eqs.~\eqref{u_n} and \eqref{Theta} at fixed $E_3$ and $\cos\theta_n$, we calculate free energy $F$(in the unit of $g_{11}$) for different values of $\eta$. In Figure~\ref{fig:plot3}, variation of $F$ with $\eta$ have been shown. For all the values of $\cos\theta_{n}$, $F$ has degenerate minima at $\eta = 0$ and $\eta = \pi$. 
	
	The solutions of $\tilde{u}_n$ and $\Theta$ in Eqs.~\eqref{u_n} and \eqref{Theta} for $\cos\theta_n = 0, \pm\frac{1}{\sqrt{2}}$ and $\pm\frac{\sqrt{3}}{2}$ at different values of $E_0/E_3$ have been shown in Fig. \ref{fig:plot4}. The profile of $\Theta(\rho)$ is found to be independent of $E_0/E_3$. The Eqs. ~\eqref{u_n} and \eqref{Theta} is same for $\eta = 0$ and $\pi$. Therefore, the solution of $\Theta$ and $\tilde{u}_{n}$ is also same for a particular value of $\cos\theta_n$ and $E_0/E_3$. We have excluded $\cos\theta_n\approx \pm 1$  because  no antiskyrmion solution is possible in these cases.
	
	In Fig. \ref{sk_antisk}, we have shown the direction of polarization vector in a plane corresponding to different functional forms of $\Phi(\phi)$. It is to be noted that Neel type outward[Fig.\ref{sk_antisk}(a)] and inward [Fig.\ref{sk_antisk}(b)] skyrmion structure exactly cancels each other if found in a system simultaneously. Similarly, Fig.\ref{sk_antisk}(d) and Fig.\ref{sk_antisk}(e) antiskyrmion structures cancels each other if  found simultaneously.

	\section{Topological Charge}\label{charge}
	The topological charge for a skyrmion or an antiskyrmion is defined as
	\begin{equation}
		Q = \frac{1}{4\pi}\int \int \bm{P} \cdot \left(\frac{\partial \bm{P}}{\partial x} \times \frac{\partial \bm{P}}{\partial y}\right)\,dxdy\,.\label{N_sk}
	\end{equation}
	where $\bm{P} = [\sin\Theta\cos\Phi,\, \sin\Theta\sin\Phi,\, \cos\Theta]$.
	The derivatives of $P$ with respect to the Cartesian coordinates are given by
	\begin{eqnarray}
		\frac{\partial \bm{P}}{\partial x} &=& \biggl(\cos\phi\cos\Theta\cos\Phi\Theta_r
		+\frac{1}{r}\sin\phi\sin\Theta\sin\Phi\Phi_{\phi}\biggr)\hat{i} \nonumber \\
		&+&\biggl(\cos\phi\cos\Theta\sin\Phi\Theta_r
		-\frac{1}{r}\sin\phi\sin\Theta\cos\Phi\Phi_{\phi}\biggr)\hat{j} \nonumber \\
		&+&\biggl(-\cos\phi\sin\Theta\Theta_r\biggr)\hat{k}\,.\label{delpdelx}
	\end{eqnarray}	
	\begin{eqnarray}
		\frac{\partial \bm{P}}{\partial y} &=& \biggl(\sin\phi\cos\Theta\cos\Phi\Theta_r
		-\frac{1}{r}\cos\phi\sin\Theta\sin\Phi\Phi_{\phi}\biggr)\hat{i} \nonumber \\
		&+&\biggl(\sin\phi\cos\Theta\sin\Phi\Theta_r
		+\frac{1}{r}\cos\phi\sin\Theta\cos\Phi\Phi_{\phi}\biggr)\hat{j}\nonumber \\
		&+&\biggl(-\sin\phi\sin\Theta\Theta_r\biggr)\hat{k}\,.\label{delpdely}
	\end{eqnarray}	
	Using Eqs.~\eqref{delpdelx} and \eqref{delpdely}, we find 
	\begin{equation}
		\bm{P} \cdot \left(\frac{\partial \bm{P}}{\partial x} \times \frac{\partial \bm{P}}{\partial y}\right) = \frac{1}{r}\sin\Theta\Theta_r\Phi_{\phi} \, .
	\end{equation}
	We thus find   
	\begin{equation}
		Q = \frac{1}{4\pi} \int_0^\infty dr \int_0^{2\pi} d\phi\, \sin\Theta \, \Theta_r \Phi_{\phi} \, .
	\end{equation}	
\end{document}